\documentclass[amsmath,amssymb,article,twocolumn,superscriptaddress,showpacs,10pt]{revtex4-1}
\usepackage{amsmath,amssymb,graphicx,bm}
\usepackage{epstopdf}
\usepackage[usenames]{color}

\begin{document}

\title[Stabilization of trapless dipolar BEC]{Stabilization of trapless dipolar Bose-Einstein condensates by temporal modulation of the contact interaction}
\author{S. Sabari$^{1,2}$ and Bishwajyoti Dey}
\affiliation{Department of Physics, SP Pune University, Pune 411007, Maharashtra, India \\ $^2$ Department of Physics, Bharathidasan University, Tiruchirappalli 620024, India}

\begin{abstract}
We theoretically study the stability of a trapless dipolar Bose-Einstein condensate (BEC) with temporal modulation of short-range contact interaction. For this aim, through both analytical and numerical methods, we solve a Gross-Pitaevskii equation with both constant and oscillatory form of short-range contact interaction along with long-range, nonlocal, dipole-dipole (DD) interaction terms. Using variational method, we discuss the stability of the trapless dipolar BEC with presence and absence of both constant and oscillatory contact interactions. We show that the oscillatory contact interaction prevents the collapse of the trapless dipolar BEC. We confirm the analytical prediction through numerical simulations. We have also studied the collective excitations in the system induced by the effective potential due to oscillating interaction. 
\end{abstract}
\maketitle

\section{Introduction}

After the experimental realization of Bose-Einstein condensates (BECs) of $^{52}$Cr~\cite{Koch:2008,Lahaye:2009}, $^{164}$Dy~\cite{Lu:2011,Youn:2010} and $^{168}$Er~\cite{Aikawa:2012} with long-range dipole-dipole (DD) interaction superposed on the short-range atomic interaction marks a major development in ultra cold quantum gases. Because of the long-range nature and  anisotropic character of the DD interaction, the dipolar BEC possesses many distinct features and new exciting phenomena such as the dependence of stability on the trap geometry~\cite{Koch:2008,Lahaye:2009}, new dispersion relations of elementary excitations~\cite{Wilson:2010,Ticknor:2011,Santos:2003}, unusual equilibrium shapes, roton-maxon character of the excitation spectrum~\cite{Santos:2003,Santos:2000,Goral:2002,Yi:2003,Ronen:2007,Boudjemaa:2013}, novel quantum phases including supersolid and checkerboard phases~\cite{Baranov:2008,Tieleman:2011,Zhou:2010}, vortices~\cite{vor1,vor2}, hidden vortices~\cite{Sabari2017}, dynamics of vortex-antivortex pairs~\cite{Sabari2018}  etc. These features arise due to the interplay between the s-wave contact interaction and the dipolar interaction. 

Tuning contact interactions using Feshbach resonance has attracted a considerable interest in the study of dipolar BECs~\cite{Goral:2000,Yi:2000,Giovanazzi:2003,Yi:2003,Giovanazzi:2006,Lin:2008,
Goral:2002,Santos:2000,Santos:2003,O'Dell:2004,Pedri:2005,Tikhonenkov:2008,Young-S:2011,Goral:2002,Cooper:2005,Zhang:2005,Rezayi:2005}. One of particular interest is macroscopically excited BEC, such as solitons. Solitons are localized waves that propagate over long distances without change in shape or attenuation. The existence of solitonic solutions is a common feature of nonlinear wave equations and solitons appear in many diverse physical systems. The theoretical description of a dilute weakly interacting dipolar BEC can be formulated by including a nonlocal DD interaction term in the Gross-Pitaevskii (GP) equation~\cite{Koch:2008,Lahaye:2009,Muruganandam:2012,Muruganandam:2011,Lahaye:2008}. The nonlinear terms in the GP equation characterized by both DD interaction and the contact interaction can support both dark and bright matter-wave solitons. In  the conventional BEC, bright matter-wave solitons form when the negative (attractive) contact interaction exactly balance with the dispersion and the attractive contact interaction~\cite{Strecker:2002,Khaykovich:1995,Khawaja:2002}. In the dipolar BECs, a nonlocal DD interaction term is involved in the nonlinear part together with the s-wave contact interaction. DD interaction has been a subject of active investigation in disparate physical systems during the past decades. DD interaction plays a crucial role in the physics of solitons and modulational instability~\cite{Krolikowski:2001,Bang:2002}. The new prospect for the formation of matter-wave bright solitons in BECs are suggested by the presence of DD interaction.  
Hence, in the presence of DD interaction, one can get a bright soliton even for positive (repulsive) contact interaction $(a > 0)$, which can be controlled by means of the Feshbach resonance with a tunable time-dependent magnetic field~\cite{Muruganandam:2012,Muruganandam:2011}. Further, in recent years, study of temporal and spatial modulated nonlinearities have attracted considerable attention in several areas, for example, nonlinear physics~\cite{Kivshar:1971}, optics~\cite{Zeng:2012,Dai:2011, Alberucci:2013} and conventional  BECs~\cite{Sabari:2010,Adhikari:2004,Saito:2003,Wu:2010}.

In conventional BECs, the periodic temporal modification of the atomic scattering length achieved by Feshbach resonance has been used to stabilize the bright solitons in higher-dimensions. Through the GP equation with constant and oscillatory part of the contact interaction, Saito and Ueda stabilized the trapless matter-wave bright solitons in 2D by temporal modulation of contact interaction~\cite{Saito:2003}, Adhikari examined the problem and stabilized the untrapped soliton in 3D and the vortex soliton in 2D by temporal modulation of contact interaction~\cite{Adhikari:2004}. Effects of the time-dependent nonlinear contact interaction on the binding energy of soliton molecules has been examined by Khawaja and Boudjemaa \cite{Khawaja:2012}. We have studied the stability of the 3D BEC with constant and oscillatory part for both the two- and three-body interactions in our previous work~\cite{Sabari:2010}. Besides, Wu et al. \cite{Wu:2010} and Wang et al. \cite{Wang:2010} discussed 2D stable solitons and vortices for BECs with spatially modulated contact interaction and a harmonic trap, respectively.

The objective of the present work is to study the significance of both constant and oscillatory part of short-range contact interaction on the stability of trapless dipolar BECs. The effective strength of the DD interactions can be controlled by adjusting the orientation of the dipoles, while the strength of the contact interactions may be effectively tuned by means of the Feshbach resonance, as shown in the condensate of $^{52}$Cr atoms. We investigate the stability of the dipolar matter-wave with both constant and oscillatory contact interaction. From our theoretical analysis, we suggest that one can increase the stability of the dipolar BEC by considering the oscillatory contact interaction. This is the main result of this paper. 

A numerical study of the time-dependent GP equation with nonlocal DD interaction term is of interest, as this can provide solutions to many stationary and time-evolution problems. In the present study, we analyze the stability of the dipolar BECs and point out that a temporal modification of the contact interaction can lead to a stabilization of the dipolar system. In addition to analytical studies, we also perform numerical verification for the stability of a dipolar BEC. In particular, by analyzing the GP equation using the variational method and direct numerical integration, we analyze the stability properties of the dipolar BEC with constant and oscillatory part of the contact interactions. Our analysis strongly suggests that the inclusion of the oscillatory contact interaction can help stabilize the dipolar BEC.

The organization of the paper is as follows. In Section~\ref{sec2}, we present a brief overview of the mean-field model. Then, we discuss the variational study of the problem and point out the possible stabilization of a trapless dipolar BEC in 2D with and without the oscillatory contact interaction in Section~\ref{sec3}. In Section~\ref{sec4}, we report the numerical results of the time-dependent GP equation through split-step Crank-Nicholson (SSCN) method. Finally, we give the concluding remarks in Section~\ref{sec5}.\\

\section{Nonlinear nonlocal model}
\label{sec2}

Consider a dipolar BEC of $N$ particles with mass $m$ and magnetic dipole moment $\mu$. At sufficiently low temperatures, the description of the ground and excited states of the condensate is described by the time-dependent, dimensionless GP equation with nonlocal DD interaction term \cite{Koch:2008,Lahaye:2009,Muruganandam:2012,Muruganandam:2011,Lahaye:2008}

\begin{align}
\begin{split}
i\hbar\frac{\partial \phi({\mathbf r},t)}{\partial t} &=\Big[-\frac{\hbar^2}{2m}\nabla^2+V({\mathbf r}) + \frac{4\pi\hbar^2a(t)N}{m}\left\vert \phi({\mathbf r},t)\right\vert^2 \\ & +N \int U_{\mathrm{dd}}({\mathbf  r}-{\mathbf r}')\left\vert\phi({\mathbf r}',t)\right\vert^2 d{\mathbf r}' \Big]\phi({\mathbf r},t), \label{eqn:dgpe}
\end{split}
\end{align}

where $V({\mathbf r}) = \frac{1}{2} m \left(\omega_x^2 x^2+\omega_y^2 y^2+ \omega_z^2 z^2 \right) \notag$, $\omega_x, \omega_y $ and $\omega_z$ are the trap frequencies, $a(t)$ is the atomic scattering length. The dipolar interaction, for magnetic dipoles, is given by $ U_{\mathrm{dd}}(\bf R)=\frac{\mu_0 \bar \mu^2}{4\pi}\frac{1-3\cos^2 \theta}{ \vert  {\bf R} \vert  ^3},$ ${\bf R= r -r'}$ determines the relative position of dipoles and $\theta$  is the angle between ${\bf R}$ and the direction of polarization $z$, $\mu_0$ is the permeability of free space and $\bar \mu$ is the dipole moment of the condensate atom. The normalization is $\int d{\bf r}\vert\phi({\mathbf r},t)\vert ^2=1.$ 
To compare the dipolar and contact interactions, often it is useful to introduce the length scale $a_{\mathrm{dd}}\equiv \mu_0 \bar \mu^2 m/(12\pi \hbar^2)$ and its experimental value for $^{52}$Cr, $^{164}$Er and $^{168}$Dy is $16a_0$, $66a_0$ and $130a_0$ respectively, with $a_0$ the Bohr radius~\cite{Koch:2008,Lahaye:2009}. 

It is convenient to use the GP equation (1) in a dimensionless form. For this purpose we make the transformation of variables as, ${\bar {\bf r}}= {\bf r}/l,{\bar {\bf R}}={\bf R}/l, \bar a(t)=a(t)/l, \bar a_{\mathrm{dd}}=a_{\mathrm{dd}}/l, \bar t=t\bar \omega$, $\bar x=x/l, \bar y=y/l, \bar z=z/l, \bar \phi=l^{3/2}\phi$, $l=\sqrt{\hbar/(m\bar \omega)}$. Eq. (\ref{eqn:dgpe})  can be rewritten (after removing the overhead bar from all the variables) as 
\begin{align}
\begin{split}
i \frac{\partial \phi({\mathbf{r}},{t})}{\partial t} & = \biggr[-\frac{1}{2}\nabla^2+ V({\mathbf{r}})+4 \pi  a(t) N \vert {\phi({\mathbf{r}},{t})} \vert^2\\
&+3N a_{\mathrm{dd}}\int \frac{1-3\cos^2\theta}{\vert \bf{R}\vert^3} \vert \phi({\mathbf{r}}',t) \vert^2 d{\mathbf{r}}'\biggr] \phi({\mathbf{r}},{t}),\label{gpe3d} 
\end{split}
\end{align}
where  $V({\mathbf{r}})=\frac{1}{2}(\gamma^2 x^2+\nu^2 y^2+\lambda^{2} z^2)$, and  $\gamma= \omega_x/\bar \omega, \nu=\omega_y/\bar \omega,\lambda =\omega_z/\bar \omega$. The reference frequency $\bar \omega $ can be taken as one of the frequencies $\omega_x, \omega_y$ or $\omega_z$ or their geometric mean $(\omega_x \omega_y \omega_z)^{1/3}$. In the following we shall use Eq. (\ref{gpe3d}) where we have  removed the `bar' from all variables. 

For an axially-symmetric ($\nu=\gamma$) disk-shaped dipolar BEC with a strong axial trap ($\lambda > \nu$, $\gamma$), we assume that the dynamics of the BEC in the axial direction is confined in the axial ground state $\phi(z)=\exp(-z^2 /2d_z^2)/(\pi d_z^2)^{1/4}, \quad d_z= \sqrt{1/\lambda},$ and we have for the wave function $\phi({\bf r})\equiv \phi(z) \times \psi(\boldsymbol{\rho},t)=\frac{1}{(\pi d_z^2)^{1/4}}\exp\left[-\frac{z^2}{2d_z^2}  \right] \psi(\boldsymbol{\rho},t)$, where  $\boldsymbol{\rho} \equiv \boldsymbol{\rho}(x,y)$, $\psi(\boldsymbol{\rho},t)$ is the effective 2D wave function for the radial dynamics and $d_z$ is the axial  harmonic oscillator length. To derive the effective 2D equation for the disk-shaped dipolar BEC, we use $\phi({\bf r})$ in Eq. (\ref{gpe3d}), multiply by the ground-state wave function $\phi(z)$ and integrate over $z$ to get the 2D equation\cite{Muruganandam:2012,Lahaye:2008}
\begin{align}
\begin{split}
&i\frac{\partial \psi(\boldsymbol{\rho},t)}{\partial t}=\Big[-\frac{\nabla_\rho^2}{2}+V(\boldsymbol{\rho})d(t)+\frac{4\pi a(t)N \vert  \psi(\boldsymbol{\rho},t)\vert^2}{\sqrt{2\pi}d_z}\\
&+ \frac{ 4\pi a_{\mathrm{dd}}N}{\sqrt{2\pi}d_z}\int \frac{d{\bf k}_\rho}{(2\pi)^2} e^{-i{\bf k}_\rho.\boldsymbol{\rho}} n({\bf k}_\rho,t)h_{2D}(\sigma)\Big] \psi(\boldsymbol{\rho},t),\label{gpe2d}
\end{split}
\end{align}
where, the parameter $d(t)$ represents the strength of the external trap which is to be reduced from 1 to 0 when the trap is switched off.   
The dipolar term has been written in the Fourier space after taking the convolution of the corresponding variables~\cite{Muruganandam:2012}, $n({\bf k}_\rho,t)=\int d\boldsymbol{\rho} e^{i{\bf k}_\rho.\boldsymbol{\rho}} \vert \psi(\boldsymbol{\rho},t) \vert^2$, $ h_{2D}= 2-3\sqrt \pi \sigma \exp(\sigma^2)\{1-\text{erf}(\sigma)\}$, $\sigma=\frac{{\bf k}_\rho d_z}{\sqrt 2}$, $\lambda=9$  and ${\bf k}_\rho=\sqrt{k_x^2+k_y^2}$. In Eq.~(\ref{gpe2d}) length is measured in units of characteristic harmonic oscillator length $l =\sqrt{\hbar/m\omega} $, angular frequency of trap in units of $\omega$, time $t$ in units of $\omega^{-1}$, and energy in units of $\hbar\omega$.

We have considered a pancake (disc) shaped condensate since azimuthal instabilities can be reduced in such condensates by choosing appropriate trap frequencies which validates the quasi-2D approximation. We have considered an axially symmetric pancake  shaped condensate in a strong axial trap  with large system parameters $\lambda = \omega_z/\bar\omega = 9$, where $\bar\omega =\omega_x=\omega_y$. We have checked from numerical simulations that stability of the dipolar BEC occurs without any axial modes or azimuthal instabilities being excited. We have repeated our simulations for higher values of the parameter $\lambda$ and got similar results.


\section{Variational Results}
\label{sec3}
In the following, to obtain the governing equations of motion of the condensate parameters, we use the variational approach with the Gaussian ansatz as a trial wavefunction for the solution of Eq. (3) where the external potential is absent [39]:

\begin{eqnarray}
\psi(\boldsymbol{\rho},t) = \frac{1}{R(t)\sqrt{\pi}}\exp{\left(-\frac{\rho^2}{2R(t)^2}+i\beta \rho^2 \right)}. \label{eqn:trial}
\end{eqnarray}
The Lagrangian density for generating Eq.~(3) with $d(t) = 0$ is,
\begin{eqnarray}
& \mathcal{L}=  \displaystyle \frac{i}{2} \left(\psi \frac{\partial\psi^*}{\partial t} -\psi^* \frac{\partial\psi}{\partial t} \right)+ \frac{\vert\nabla_{\rho} \psi \vert^2}{2}  + \frac{2\pi Na(t)}{\sqrt{2\pi}d_z} \vert  \psi\vert^4  \nonumber \\
&\,\displaystyle+\frac{2\pi Na(t)}{\sqrt{2\pi}d_z}\vert  \psi\vert^2  \int\frac{d{\bf k}_{\rho}}{(2\pi)^2}e^{i{\bf k}_{\rho}.\boldsymbol{\rho}} n({\bf k}_\rho,t)h_{2D}(\sigma)\Bigg), \label{Jac3}
\end{eqnarray}

The trial wave function (\ref{eqn:trial}) is substituted in the Lagrangian density (\ref{Jac3}) and the effective Lagrangian per particle is calculated by integrating the Lagrangian density as 
\begin{align}
\begin{split}
 \it{L_{eff}} = &\, 2\,R(t)^2\,\beta(t)^2+\frac{N a(t)}{\sqrt{2\pi} d_z \, R(t)^2}+\frac {1}{2R(t)^2} \nonumber \\
&+\, R(t)^2\,\dot{\beta}(t)-\frac{a_{dd}\,\eta(\xi)}{\sqrt{2\pi} d_z \, R(t)^2}, \label{Jac3a}
\end{split} 
\end{align}

 \noindent with
 \begin{equation} \eta(\xi)=\frac{1+2\xi^2-3\xi^2 d(\xi)}{(1-\xi^2)}, \,\,\,d(\xi)=\frac{atanh\sqrt{1-\xi^2}}{\sqrt{1-\xi^2}} \nonumber \\\end{equation} 
 and $\xi=R(t)/d_z$. The Euler-Lagrangian equations for the variational parameters $R(t)$ and $\beta(t)$ are obtained from the effective Lagrangian in a standard fashion as

\begin{eqnarray} 
\frac{\partial R(t)}{\partial t} & = &2R(t)\beta(t),\label{Jac3b}\\
\frac{\partial \beta(t)}{\partial t}& = &\frac {1}{2R(t)^4}+\frac{N(a(t)-a_{dd}\,\eta(\xi))}{\sqrt{2\pi} d_z \, R(t)^4}-2\beta(t)^2, \label{Jac3c}
\end{eqnarray}

By combining equations (\ref{Jac3b}) and (\ref{Jac3c}), we get the following
second-order differential equation for the evolution of the
width $R(t)$:
\begin{eqnarray}
 \frac{\partial^2 R(t)}{\partial t^2} & = & \frac {1}{R(t)^3}+\frac{2 Na(t)}{\sqrt{2\pi} d_z \, R(t)^3}-\frac{Na_{dd}\,\Lambda(\xi)}{\sqrt{2\pi} d_z \, R(t)^3}, \label{Jac4} \end{eqnarray}
where $\Lambda(\xi)=2-7\xi^2-4\xi^4+9\xi^4 d(\xi)/(1-\xi^2)^2$.
 
Since, we consider a periodic modulation of the s-wave interaction of the form $a(t) = \epsilon_0 + \epsilon_1 \sin{(\Omega t)}$ on the stability of dipolar BEC, where $\epsilon_0$, and $\epsilon_1 $ are the amplitudes of constant and oscillating part of s-wave contact interaction, respectively, a Kapitza averaging scheme can be used to treat these oscillatory terms~\cite{Landau:1960}. Such a modulation of the contact interaction is possible by manipulating an external magnetic or optical field near a Feshbach resonance~\cite{Kivshar:1971,Zeng:2012,Adhikari:2004,Dai:2011,Sabari:2010,Saito:2003,
Wu:2010,Wang:2010}. After including the oscillating nonlinearity in the contact interaction part, we get the following second-order differential equation for the evolution of the width for radial coordinates \cite{Muruganandam:2012},

\begin{eqnarray}
\frac{\partial^2 R(t)}{\partial t^2} & = & \frac {1}{R(t)^3}+ \frac{2N [\epsilon_0+\epsilon_1\sin(\Omega t)]-Na_{dd}\,\Lambda(\xi)}{\sqrt{2\pi} d_z \, R(t)^3}, \,\,\, \label{Jac6}
\end{eqnarray}

Now $R$ can be separated into a slowly varying part $R_0$ and a rapidly varying part $R_1$ by $R=R_0+R_1$. When $ \Omega\ \gg 1 $, $R_1$ becomes of the order of $\Omega^{-2}$. Keeping the terms of the order of up to $\Omega^{-2}$ in $R_1$, we obtain the following equations of motion for $R_0$ and $R_1$~\cite{Landau:1960},

\begin{eqnarray} 
\frac{\partial^2 R_1}{\partial t^2} & = &\frac{2N\epsilon_1\sin(\Omega t)}{\sqrt{2\pi} d_z \,R_0^3},\label{Jac9}\\
\frac{\partial^2 R_0}{\partial t^2}& = &\frac {1}{R_0^3} + \frac{N[2\epsilon_0-a_{dd}\,\Lambda(\xi_0)]}{\sqrt{2\pi} d_z \,R_0^3}  -\frac{6N\epsilon_1 \langle R_1\sin(\Omega t)\rangle}{\sqrt{2\pi} d_z \,R_0^4}, \nonumber \\ \label{Jac10}
\end{eqnarray}
where $\langle \cdots \rangle$ denotes the time average over the rapid oscillation. From Eq.~(\ref{Jac9}), using the solution $R_1=-2N \epsilon_1 \sin(\Omega t) /[\sqrt{2\pi} d_z \, \, \Omega^2 R_0^3]$ and substituting it into Eq.~(\ref{Jac10}), we obtain the following equation of motion for the slowly varying part,
\begin{equation}
\frac{d^2 R_0}{d t^2}=\frac {1}{R_0^3} + \frac{N[2\epsilon_0-a_{dd}\,\Lambda(\xi_0)]}{\sqrt{2\pi} d_z \,R_0^3}+\frac{6N^2\epsilon_1^2}{2\pi d_z^2\, \Omega^2\,R_0^7}. \label{Jac11}
\end{equation}
The variational approximation suggests that the effect of the DD interaction is to reduce the constant contact interaction for $a_{dd} >0$. Immediately, one can conclude that the system effectively becomes attractive for $a_{dd} > \epsilon_0$. So one can have the formation of bright soliton even for positive (repulsive) scattering length, provided that $a_{dd} > \epsilon_0$. 

\begin{figure}[h!]
\begin{center}
\includegraphics[width=0.8\linewidth]{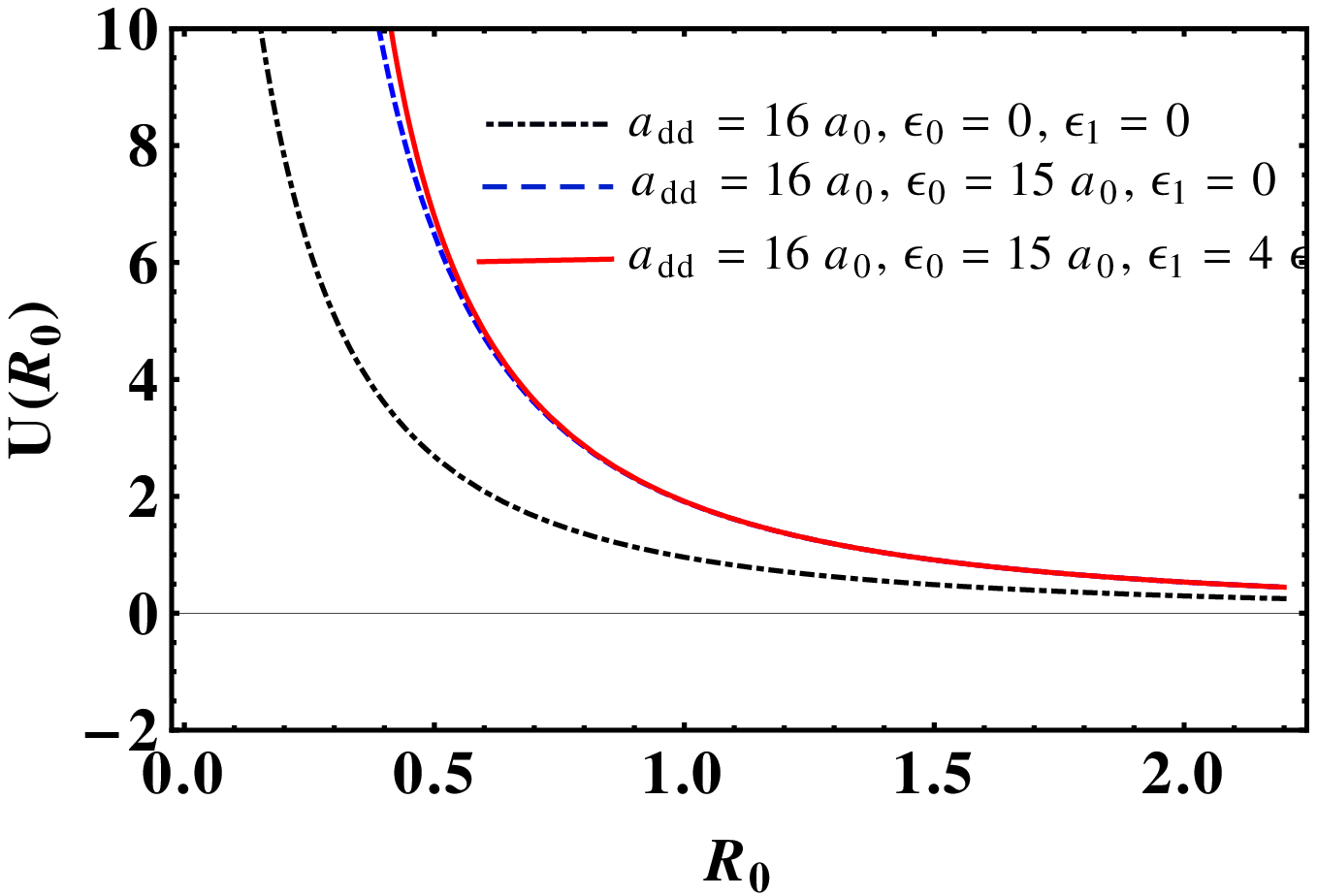}\\
\includegraphics[width=0.8\linewidth]{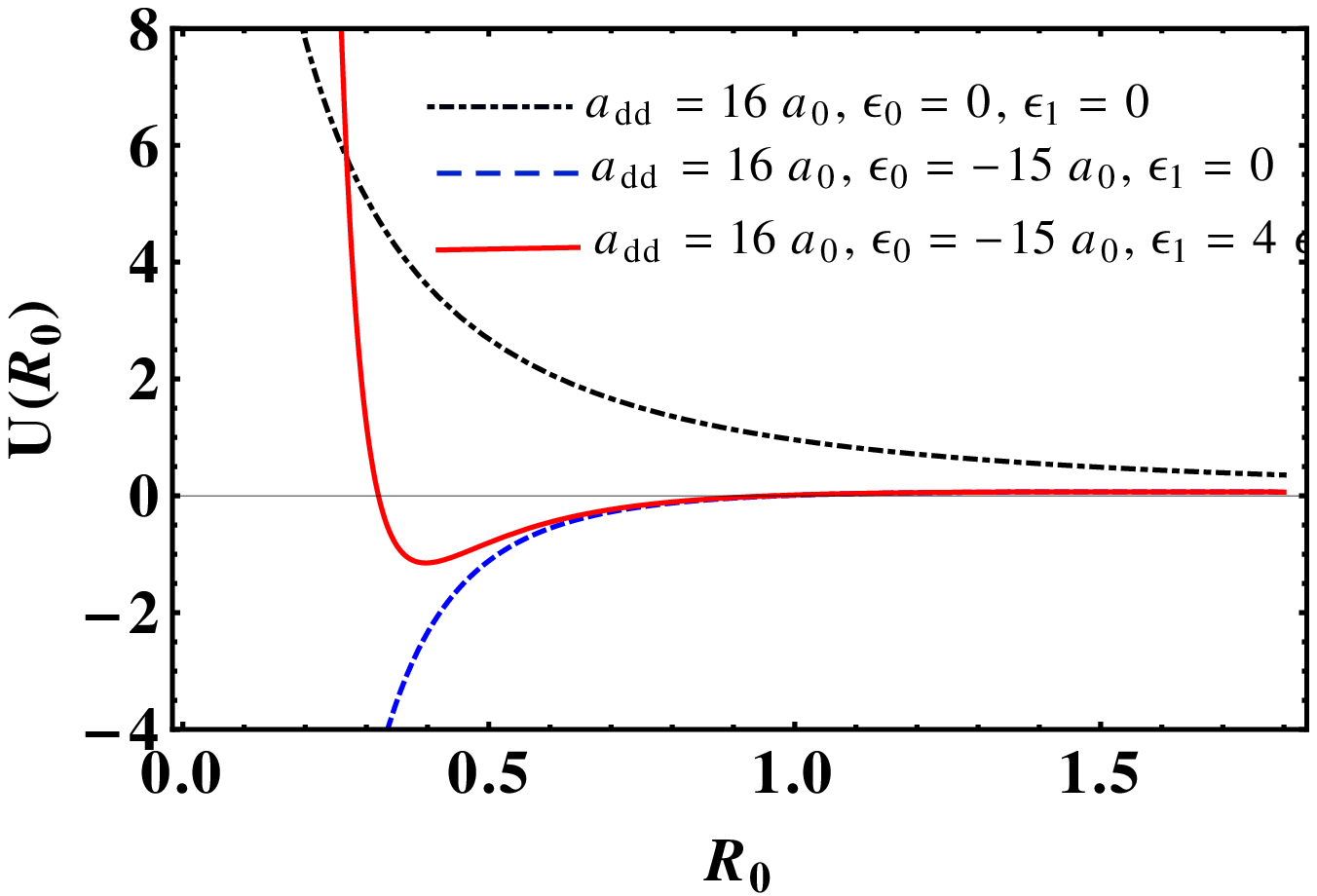}
\end{center}
\caption{The effective potential using Eq. (\ref{Jac11}) showing the stability properties of the trapless dipolar condensate.}
\label{f1}
\end{figure}
\noindent 
Eq. (12) can be written as $\frac{d^2 R_0}{d t^2}= - \frac {\partial U(R_0)}{\partial R_0}$ where the effective potential $U(R_0)$ is given by,
\begin{equation}
 U(R_0) = \frac {1}{2\,R_0^2} + \frac{N[2\epsilon_0-a_{dd}\,\eta(\xi_0)]}{2\sqrt{2\pi} d_z \,R_0^2}+\frac{N^2\epsilon_1^2}{2\pi d_z^2\,\Omega^2 \,R_0^6}. \label{Jac13}
\end{equation}
\noindent where $ \eta(\xi_0)=[1+2\xi_0^2-3\xi_0^2d(\xi_0)]/ (1-\xi_0^2)$ and $\xi_0=R_0/d_z$. Now, we can analyze the nature of the effective potential $U(R_0)$ versus $R_0$ in the presence and absence of $a_{dd}$ and $\epsilon_1$.

\begin{figure}[h!]
\begin{center}
\includegraphics[width=0.8\linewidth]{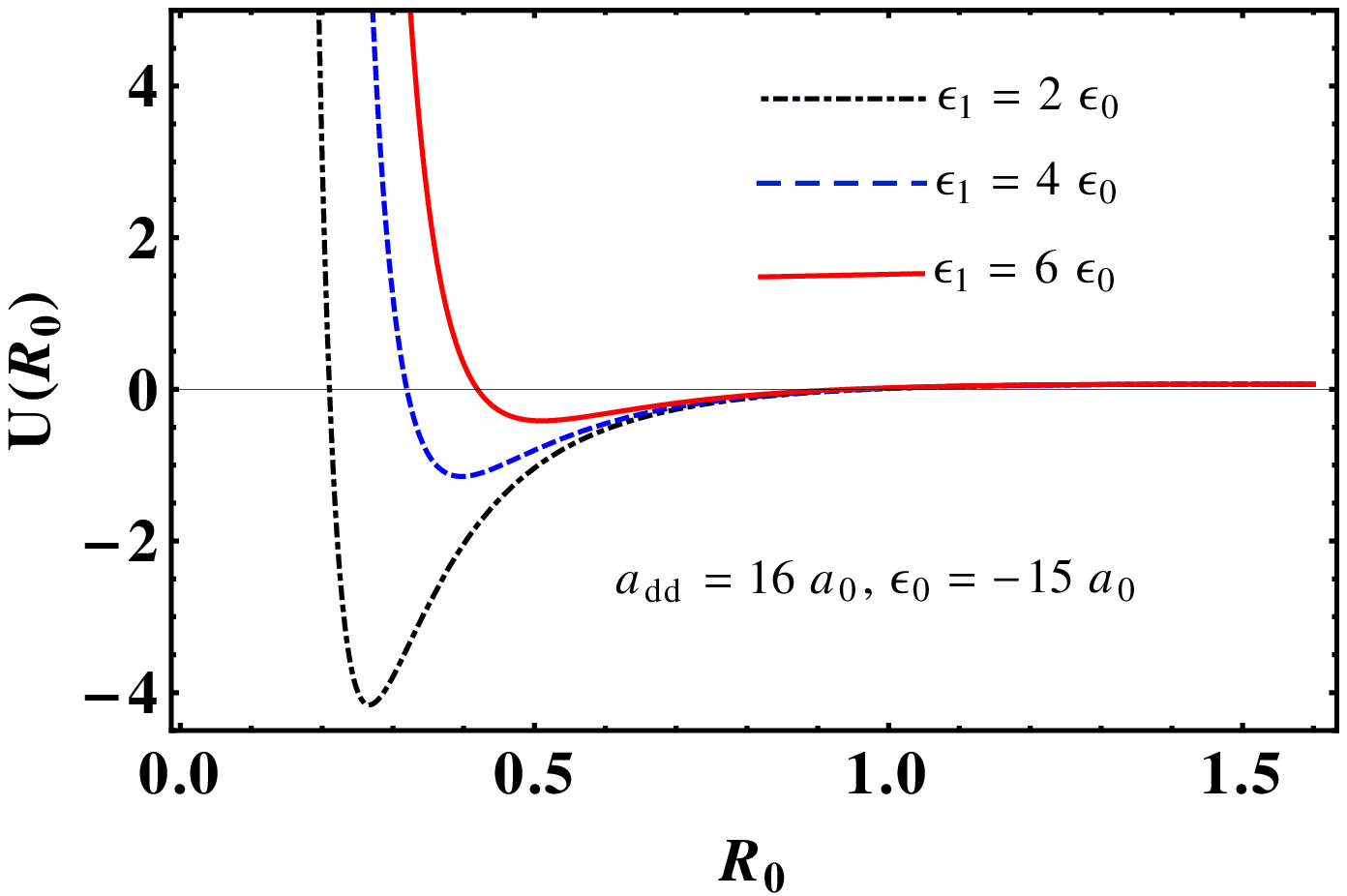}\\
\includegraphics[width=0.8\linewidth]{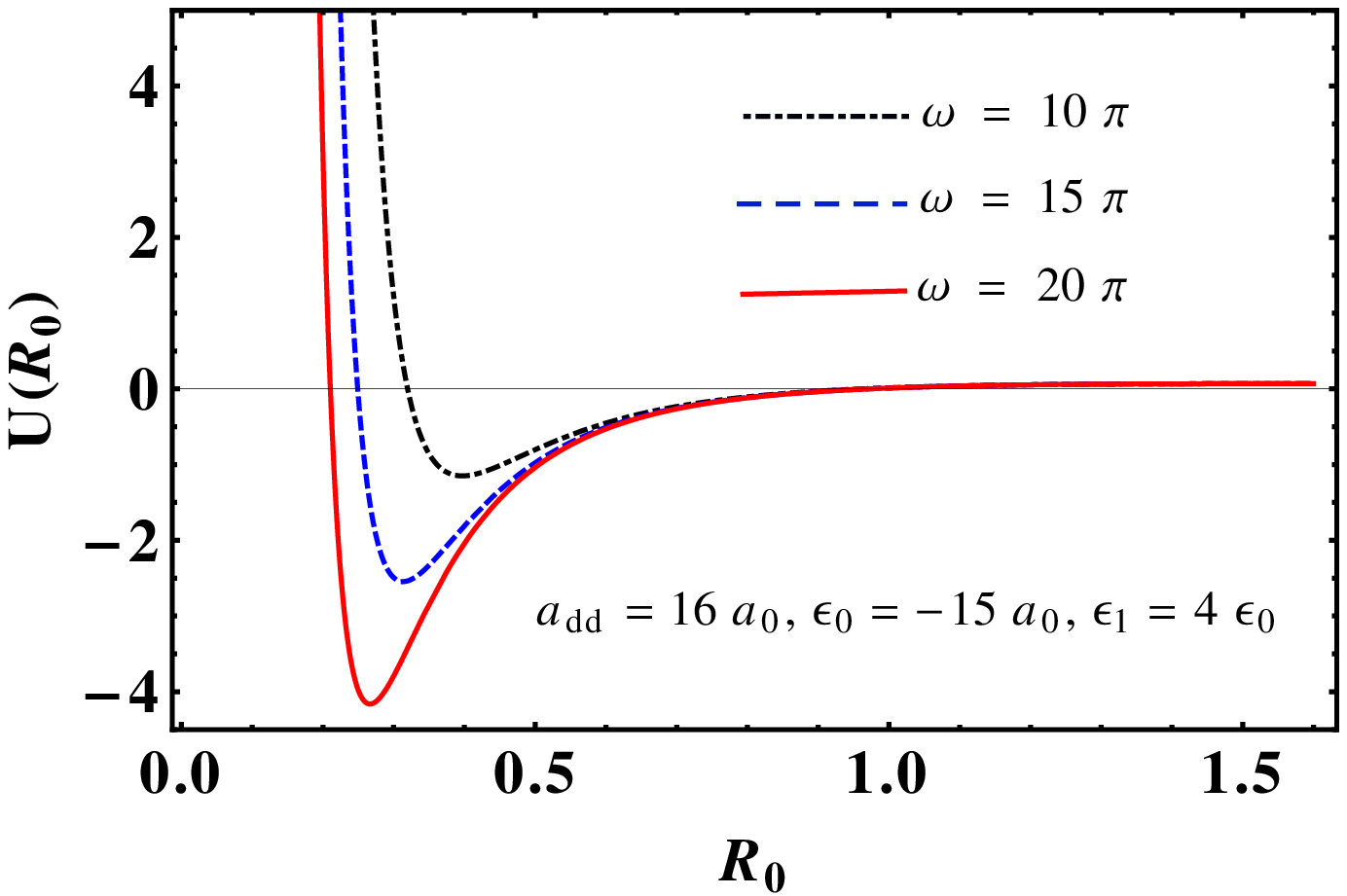}
\end{center}
\caption{(Upper panel)Potential curves for varying values of the amplitude of oscillating nonlinearity. (lower panel) Potential curves for varying values of frequency of oscillation of the two-body interaction.}
\label{f3}
\end{figure}

In Fig.~(\ref{f1}), we show the stability properties of the $^{52}$Cr condensate in the absence of external trap potential in panel (a) for repulsive two-body ($\epsilon_0=15 a_0$) and panel (b) for attractive two-body ($\epsilon_0=-15 a_0$) interactions. And the other parameters are taken as, $a_{dd}=16 a_0$, $\epsilon_1=4\epsilon_0$, $\Omega = 10\pi$, and $N=1000$. In both panels, dash-dotted curve represents the potential in the presence of dipolar interaction alone, dashed curve for dipolar interaction with constant part of contact interaction, and continuous curve for dipolar with both constant and oscillatory part of the contact interactions. In top panel, we observe no negative value of the potential for all three cases. Moreover, there is no trapping potential to prevent the expansion of the width of the condensate. 

But in the lower panel, we observe negative potential value for condensate with DD interaction in-addition to both constant and oscillatory contact interaction (continuous curve). Here, the attraction due to constant part of the two-body and DD interactions is balanced by the oscillatory part of the two-body interaction. The dashed curve corresponds to collapse of BEC due to strong attractive contact two-body interaction. 

\begin{table}[!ht]
\caption{Stability properties of trapless dipolar BEC by variational method}
\label{table1}
\begin{center}
\begin{tabular}{|c|c|c|c|c|c|c||}
\hline \hline
& & & & & U($R_0$)   &$R_0$ at  U($R_0$) \\
$\epsilon_0$  &$a_{dd}$ &$\epsilon_1$  &$\Omega$ & minimum & at minimum & minimum  \\
\hline \hline
 0       &16$a_0$ &0             &$-$     &No  &$-$      & $-$ \\ \hline
 15$a_0$ &16$a_0$ &0             &$-$     &No  &$-$      & $-$ \\
-15$a_0$ &16$a_0$ &0             &$-$     &No  &$-$      & $-$ \\ 
 15$a_0$ &16$a_0$ &4$\epsilon_0$ &10$\pi$ &No  &$-$      & $-$ \\
-15$a_0$ &16$a_0$ &4$\epsilon_0$ &10$\pi$ &Yes &$-0.258$ &$0.560$ \\ \hline
-20$a_0$ &16$a_0$ &4$\epsilon_0$ &10$\pi$ &Yes &$-0.900$ &$0.567$ \\
-25$a_0$ &16$a_0$ &4$\epsilon_0$ &10$\pi$ &Yes &$-1.453$ &$0.582$ \\ \hline
-15$a_0$ &16$a_0$ &$\epsilon_0$  &10$\pi$ &Yes &$-5.237$ &$0.246$ \\
-15$a_0$ &16$a_0$ &2$\epsilon_0$ &10$\pi$ &Yes &$-1.550$ &$0.365$ \\
-15$a_0$ &16$a_0$ &4$\epsilon_0$ &10$\pi$ &Yes &$-0.259$ &$0.560$ \\
-15$a_0$ &16$a_0$ &6$\epsilon_0$ &10$\pi$ &Yes &$-0.003$ &$0.749$ \\
-15$a_0$ &16$a_0$ &7$\epsilon_0$ &10$\pi$ &No  &$-$      &$-$     \\ \hline
-15$a_0$ &16$a_0$ &4$\epsilon_0$ &15$\pi$ &Yes &$-0.900$ &$0.567$ \\
-15$a_0$ &16$a_0$ &4$\epsilon_0$ &20$\pi$ &Yes &$-1.453$ &$0.582$ \\
\hline\hline
\end{tabular}
\end{center}
\end{table}

In Fig.~(\ref{f3}), the effect of the amplitude of the oscillatory contact interaction ($\epsilon_1$) is depicted for with and without the function $\epsilon_0$. It is evident from two panels, that there is an enhancement of the stability region of trapless dipolar BEC due to the inclusion of oscillatory contact interaction in addition to the constant part of the contact interaction. In the present work, we have used the relation $\epsilon_1/\epsilon_0=4$ which is already used in \cite{Sabari:2010,Adhikari:2004,Saito:2003}. Also, if we increase the amplitude of the oscillatory part alone, $\epsilon_1/\epsilon_0>4$, it decreases the depth of the minimum in the effective potential and the trapless system will become unstable due to more oscillation when compare with the attraction due to both DD interaction ($a_{dd}$) and constant part of two-body interaction ($\epsilon_0$). But, if we decrease the amplitude of the oscillatory part alone, $\epsilon_1/\epsilon_0<4$, it increases the depth of the minimum in the effective potential and the trapless system will become more stable. This is illustrated in the upper panel in Fig.~(\ref{f3}). Moreover, the trapless condensate is more stable for higher frequency of oscillation $\Omega$. In Table 1, we systematically present the stability properties of trapless dipolar BEC with the effect of oscillatory two-body contact interaction. In the following, we confirm these predictions using direct numerical integration of the governing equation.\\

\section{Numerical Results}
\label{sec4}
We solve the GP Eq.~\eqref{gpe2d} by employing real-time propagation with split-step Crank-Nicolson method applied to the diffraction operator~\cite{Muruganandam:2009,RKK:2015}. The DD interaction is evaluated by fast Fourier transform~\cite{Goral:2002}. The typical discretized space and time steps for the numerical grid is $0.05$ and $0.005$, respectively. In the numerical simulation, it is important to remove the harmonic trap while increasing the nonlinearity for obtaining the stability. Otherwise the oscillations that arise due to sudden removal of trap may lead to collapse due to attraction. In the course of time iteration, the coefficients of the nonlinear terms are increased from $0$ at each time step as $g(t) = f(t) g_f  \{a_1-b_1\sin[\,\Omega\, t]\}$, with $f(t)=t/\tau$ for $0 \leq t \leq \tau$, $f(t)=1$ for $t > \tau$ and $g_f=4\pi N a$. At the same time the trap is removed by changing $d(t)$ from $1$ to $0$ by $d(t)=1-f(t)$. 
During this process, the harmonic trap is removed, and after the $g_f$ is attained at time $\tau$, the periodically oscillating nonlinearity $g(t)=g_f[a_1-b_1 \sin \Omega t  ]$ is effected for $t> \tau$~\cite{Sabari:2010,Adhikari:2004,Saito:2003}. In the following, we present results for $^{52}$Cr atoms which has a moderate dipole moment with $a_{dd}$ = $16a_0$~\cite{Koch:2008,Lahaye:2009}. 
\begin{figure}[h!]
\begin{center}
 \includegraphics[width=0.95\linewidth]{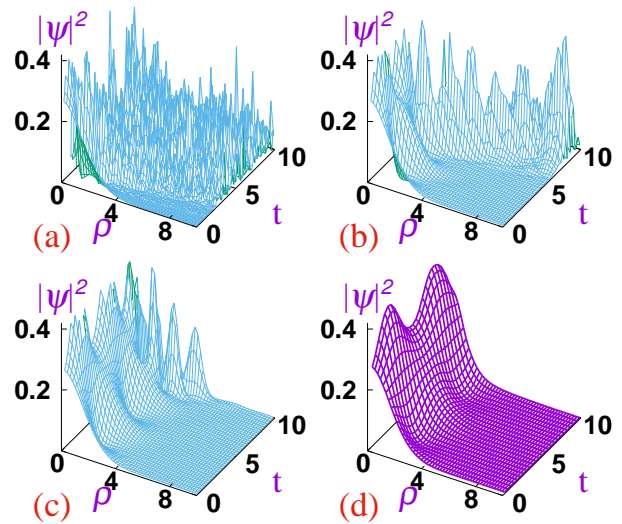}
\end{center}
\caption{Density profiles for trapless dipolar BEC with different values  constant part of the two-body interaction. Panel (a) $\epsilon_0=-15a_0$, $\epsilon_1=0$, (b) $\epsilon_0=-20a_0$, $\epsilon_1=4 \epsilon_0$, (c) $\epsilon_0=-25a_0$, $\epsilon_1=4 \epsilon_0$ and (d) $\epsilon_0=-28a_0$, $\epsilon_1=4 \epsilon_0$. Other parameters are $a_{dd}=16 a_0$, $N=1000$ and $\Omega=10\pi$.}
\label{fig:nu1}
\end{figure}
We consider different dynamical regimes wherein we alternatively study the effects of inclusion of the time-dependent periodic two-body interaction as well as the DD interaction so as to understand their effects on the system dynamics. 

\begin{figure}[h!]
\begin{center}
\includegraphics[width=0.95\linewidth]{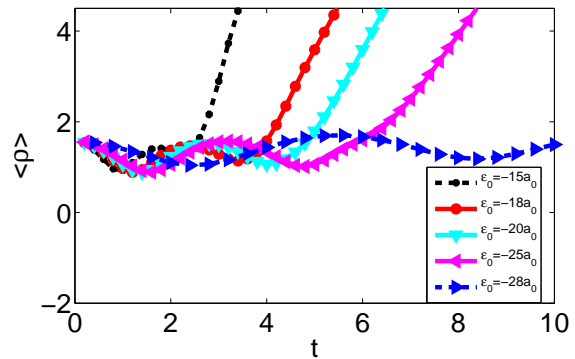}
\end{center}
\caption{Plot of the root mean squared distance $<\rho>$ rms as a function of time t for different values  constant part of the two-body interaction.}
\label{fig:nu1rms}
\end{figure}
In Fig.~(\ref{fig:nu1}), we show the dynamical stabilization of the trapless dipolar BEC. It is already known that the repulsive case the trapless condensate expands in time on the other hand for attractive case it collapse in time. But in the presence of oscillatory interaction ($\epsilon_{1}$), the stability of the trapless condensate is progressively increasing with increasing constant part of the two-body interaction along with its oscillatory part ($\epsilon_{1}$). This is clearly explain in Fig.~\ref{fig:nu1}. Here the frequency of oscillation of the time-periodic term is kept constant at $\Omega=10\pi$. In panel (a) density profile shows the dynamics for $\epsilon_0=-15a_0$, $\epsilon_1=0$. Here, density get collapses due to strong attraction by the combination of both DD interaction and the constant two-body interaction. But, if we include the oscillatory contact interaction ($\epsilon_1=4 \epsilon_0$) the system become stable upto $t=5$ time units in panel (b). Further, if we increase $\epsilon_0$ to in (c) $\epsilon_0=-25a_0$ and (d) $\epsilon_0=-28a_0$, the becomes stable upto $t=6$ and $t=10$ time units, respectively. 

In Fig.~\ref{fig:nu1rms}, we show the root mean squared distance $<\rho>_{rms}$ of the trapless dipolar BEC. For $\epsilon_1=4 \epsilon_0$, the $<\rho>_{rms}$ of the condensate is stable upto $t=2$ time units. But, if we increase $\epsilon_0$ to $\epsilon_0=-18a_0$, $\epsilon_0=-20a_0$, $\epsilon_0=-25a_0$ and $\epsilon_0=-28a_0$, the $<\rho>_{rms}$ of the condensate is becomes stable upto $t=4$, $t=5$, $t=6$ and $t=10$ time units, respectively. The oscillation and exponential growth of the curves is for stable and collapse of the system, respectively. From Figs.~(\ref{fig:nu1}) and (\ref{fig:nu1rms}), it is clear that oscillating contact interaction can help in stabilizing the trapless dipolar BEC, because the effect of the oscillatory term is to induce an additional potential due to Kapitza averaging \cite{Sabari:2010,Adhikari:2004,Saito:2003} the profile and magnitude of which depends on the frequency of oscillation.

\begin{table}[!ht]
\caption{Stability properties of trapless dipolar BEC (comparison)}
\label{table2}
\begin{center}
\begin{tabular}{|c|c|c|c|c|c||}
\hline \hline
& & & & U($R_0$)Min. & Inference    \\
$\epsilon_0$  &$a_{dd}$ &$\epsilon_1$  &$\Omega$ & (analytic) & (numerics) \\
\hline \hline
 15$a_0$ &16$a_0$ &0             &$-$     &No  &Unstable (Expand)      \\
-15$a_0$ &16$a_0$ &0             &$-$     &No  &Unstable (Collapse)     \\ 
 15$a_0$ &16$a_0$ &4$\epsilon_0$ &10$\pi$ &No  &Unstable (Expand)    \\
-15$a_0$ &16$a_0$ &4$\epsilon_0$ &10$\pi$ &Yes &Stable (upto 2 time units)\\ 
-18$a_0$ &16$a_0$ &4$\epsilon_0$ &10$\pi$ &Yes &Stable (upto 4 time units)\\ 
-20$a_0$ &16$a_0$ &4$\epsilon_0$ &10$\pi$ &Yes &Stable  (upto 5 time units)\\
-25$a_0$ &16$a_0$ &4$\epsilon_0$ &10$\pi$ &Yes &Stable  (upto 6 time units)\\
-28$a_0$ &16$a_0$ &4$\epsilon_0$ &10$\pi$ &Yes &Stable  (upto 10 time units)\\ 
-20$a_0$ &16$a_0$ &$\epsilon_0$  &15$\pi$ &Yes &Stable (upto 18 time units) \\ 
-20$a_0$ &16$a_0$ &$\epsilon_0$  &20$\pi$ &Yes &Stable (upto 20 time units) \\ 
-20$a_0$ &16$a_0$ &$\epsilon_0$  &24$\pi$ &Yes &Stable (upto 50 time units) \\
\hline\hline
\end{tabular}
\end{center}
\end{table} 

\begin{figure}[h!]
\begin{center}
 \includegraphics[width=0.95\linewidth]{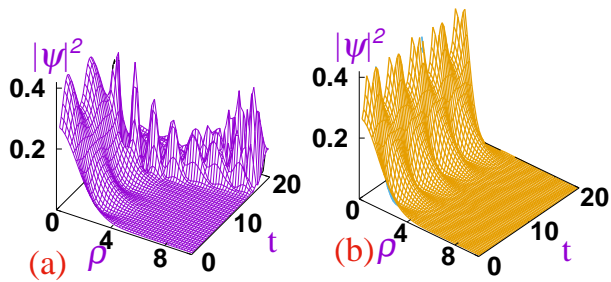}
\end{center}
\caption{Density profiles for (a) $\epsilon_0=-28a_0$, $\Omega=10\pi$ and (b) $\epsilon_0=-20a_0$, $\Omega=20\pi$. Other parameters are $a_{dd}=16 a_0$, $N=1000$ and $\epsilon_1=4\epsilon_0$.}
\label{fig:nu2}
\end{figure}

\begin{figure}[h!]
\begin{center}
\includegraphics[width=0.95\linewidth]{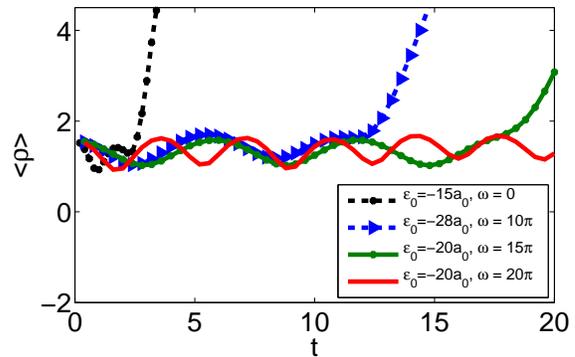}
\end{center}
\caption{Plot of the root mean squared distance $<\rho>$ rms as a function of time t.}
\label{fig:nu2rms}
\end{figure}

Now we illustrate the effect of varying frequency of oscillation of the time-periodic oscillatory contact interaction. The results are depicted in Fig.~\ref{fig:nu2} for increasing values of the frequency of oscillation $\Omega$. As can be seen, increase in oscillation frequency further helps in the stabilization.
In Fig.~\ref{fig:nu2rms}, we show the root mean squared distance $<\rho>_{rms}$ for increasing values of the frequency of oscillation $\Omega$. From Fig. 4 and Fig. 6 we can see that the oscillation in $<\rho>$ persists only for a particular duration of time depending on the system parameters. From Eq. (13) we can see that when $\epsilon_1$ is small or when $|\epsilon_0|$ is large, then the effective force is not sufficient to prevent the atoms from collecting at the center. As a result the peak density $|\psi(\rho = 0)|^2$ grows and the condensate width decreases. The condensate then expands due to repulsive first and the last terms in the effective potential (Eq. (13)). Subsequently most of the expanded atoms collects at the center due to the attractive terms (second term) in the effective potential (Eq. (13)). This process of oscillation (contraction and expansion) of the condensate goes on for some time. Since the system is trapless, after each expansion some of the atoms scattered with high energy cannot return to the center. The oscillation decays and the condensate starts to expand as shown in Fig. 4 and Fig. 6. From Figs.~(\ref{fig:nu1} - \ref{fig:nu2rms}), it is clear that oscillating contact interaction can help in stabilizing the trapless dipolar BEC, because the effect of the oscillatory term is to induce an additional potential due to Kapitza averaging \cite{Sabari:2010,Adhikari:2004,Saito:2003} the profile and magnitude of which depends on the frequency of oscillation. A comparison between the analytical and numerical results for different parameters values is summarised in Table II showing good agreement.\\
\section{Collective excitations} \label{sec5}
It is well known that collective modes can be induced in the condensate by several means, such as, rotating the condensate, modulation of the external trapping potential, modulation of the $s$-wave  scattering length etc. It has been shown that the excitations of low lying collective modes, like the breathing mode, can be induced by harmonic modulation of the $s$-wave  scattering length \cite{Vidanovic:2011,Saito:2003}. Bismut et. al.  \cite{Bismut:2010} measured the effect of dipole-dipole interactions on the frequency of a collective mode of a Cr BEC. Recently we have studied the
hydrodynamics of collective excitations like quantized vortices and solitons in a dipolar Bose-Einstein condensate induced by an oscillating trapping potential \cite{Sabari:2017}.\\
In the present case, we study the collective excitation of the condensate using variational method and numerical simulations.  From Eq. (13)  we obtain the minimum of the effective potential $U(R_0)$ at
\begin{equation}
R^4_{{\rm min}} = -\frac{6N^2\epsilon_1^2}{d_z^2\Omega^2[2\pi + 
{\sqrt \frac{\pi}{2}}N(2\epsilon_0  -  a_{dd}\eta(\xi_0))]}
\end{equation}
To obtain the frequency of the breathing mode (small oscillation) we linearlize Eq. (12) around the minimum of the effective potential $U(R_0)$ (Eq. (13)) \cite{Saito:2003}. For this we expand the effective potential around the minimum of the potential by Taylor series keeping only upto quadratic term in the expansion. The frequency of the small oscillation or the breathing mode around the minimum  as obtained variationally is given by  
\begin{figure}[h!]
\begin{center}
\includegraphics[width=0.95\linewidth]{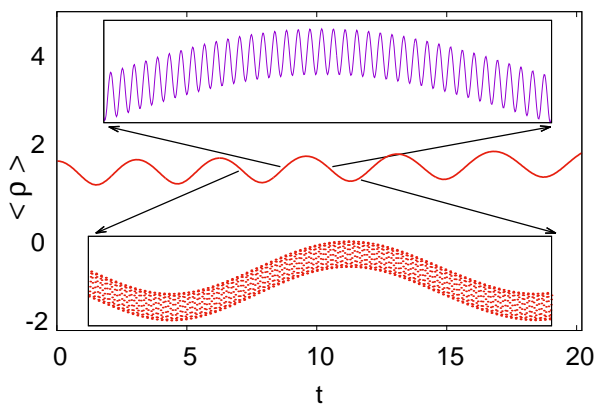}
\end{center}
\caption{Plot of the root mean squared distance $<\rho>$ rms as a function of time t for $\epsilon_0=-20a_0$, $\Omega=20\pi$, $a_{dd}=16 a_0$, $N=1000$ and $\epsilon_1=4\epsilon_0$.}
\label{fig:brfreq}
\end{figure}
\begin{equation}
\omega_{\rm br}^2 = \frac{d_z^2\Omega^2[2\pi + {\sqrt \frac{\pi}{2}}N(2\epsilon_0 - a_{dd}\eta(\xi_0))]^2}
{3\pi  N^2\epsilon_1^2}
\end{equation}
Fig. ~\ref{fig:brfreq} shows the breathing mode as obtained from the numerical simulations. The figure shows both the rapid oscillation part and a slow, smoothly varying breathing mode caused by the effective potential due to the oscillating interaction.
\section{Conclusion} \label{sec6}
In conclusion, we have stabilized the trapless dipolar Bose-Einstein condensate by considering constant and oscillatory part of the short-range contact interaction. The effect of the oscillatory nonlinear term in the short-range contact interaction is to provide an additional confining potential which helps in the stabilization of the trapless dipolar BEC. For this aim, we first performed a variational analysis on the governing equation and obtained the equations of motion. Using this we derived the effective potential which is experienced by the system. A minimum in the potential signifies a possible stable state. Our study have shown that stability of the dipolar BEC can be increased by considering the oscillatory contact interaction. We have shown that the dipolar BEC can be stabilized over various length of time for appropriate choice of the system parameters. To further prove this point, we performed direct dynamical evolution of the condensate using the harmonic oscillator solution to understand the stability properties. As from variational analysis, numerically also we conclude that the trapless dipolar BEC is stabilized and also the stability of the trapless dipolar condensate is highly enhanced by the time-periodic contact interaction in addition to the constant part of the contact interaction. Even though the periodic temporal modification of the atomic scattering length achieved by Feshbach resonance has been used to stabilize the conventional BECs, to the best of our knowledge, such experimental or theoretical studies has not yet been reported for the dipolar condensate. Such stable dipolar BECs can be used for different applications which require condensates to remain stable over large time scales. Our predictions of dynamically stabilizing a dipolar condensate by temporal modulations of the two-body interaction can be tested in experiments with pancake shaped condensate. We have also studied the collective excitations in the system induced by the effective potential due to oscillating interaction. 

\section*{Acknowledgements}
\noindent SS wishes to thank UGC for Dr. D.S. Kothari Post Doctoral Fellowship (No.F.4-2/2006 (BSR)/PH/14-15/0046). The work is partially supported by DST-SERB for Post-Doc through National Post Doctoral Fellowship (NPDF) Scheme (Grant No. PDF/2016/004106). BD acknowledge the Science and Engineering Research Board (SERB), Government of India, for funding under Grant No. EMR/2016/002627.

\end{document}